\documentclass[onecolumn,superscriptaddress,showpacs,showkeys,preprint]{revtex4}
\usepackage{graphicx}
\usepackage[latin1]{inputenc}
\usepackage[english]{babel}
\usepackage{amssymb}
\usepackage{amsmath}

\newcommand{\eg}{{\it e.g.}}
\newcommand{\ie}{{\it i.e.}}
\newcommand{\cf}{{\it cf.}}
\newcommand{\GeV}{\;\mathrm{GeV}}
\newcommand{\MeV}{\;\mathrm{MeV}}
\newcommand{\nutev}{NuTeV}
\newcommand{\pythia}{{PYTHIA}}
\newcommand{\erf}{\mathrm{erf}}
\begin{document}

\selectlanguage{english}

\title{Quark asymmetries in the proton
from a model\\ for parton densities}
\date{\today}
\author{J.~Alwall}
\email[E-mail: ]{johan.alwall@tsl.uu.se}
\affiliation{High Energy Physics, Uppsala University, Box 535, S-75121 Uppsala, Sweden}
\author{G.~Ingelman}
\email[E-mail: ]{gunnar.ingelman@tsl.uu.se}
\affiliation{High Energy Physics, Uppsala University, Box 535, S-75121 Uppsala, Sweden}

\begin{abstract}
Based on quantum fluctuations in momentum and of the proton into meson-baryon pairs, we develop a physical model for the non-perturbative $x$-shape of parton density functions in the proton. The model describes the proton structure function and gives a natural explanation of observed quark asymmetries, such as the difference between the anti-up and anti-down sea quark distributions and between the up and down valence distributions. An asymmetry in the momentum distribution of strange and anti-strange quarks in the nucleon is found to reduce the NuTeV anomaly to a level which does not give a significant indication of physics beyond the standard model. 
\end{abstract}

\preprint{TSL/ISV-2005-0288}

\maketitle

%%%%%%%%%%%%%%%%%%%%%%
\section{Introduction}
%%%%%%%%%%%%%%%%%%%%%%

The parton distributions in hadrons play a very important role in
particle physics. Cross-sections for hard processes involving incoming
hadrons can, based on the factorization theorems of
QCD~\cite{factorisation-theorem}, be calculated as a convolution of
the parton distributions with parton level cross-sections calculated
using perturbation theory. The parton distributions are universal in
the sense that each hadron has a unique parton structure which can be
used to calculate all hard processes involving that hadron. The normal
interpretation of a parton distribution $f_i(x,Q^2)$ is as the
probability to find a parton $i$ (quark of some flavor or gluon) with
a fraction $x$ of the hadron momentum when probed by the momentum
transfer $Q^2$. The $Q^2$-dependence is very successfully described in
perturbative QCD (PQCD) by the $\log Q^2$ evolution equations~\cite{DGLAP}, which means that, given the input distributions in $x$ at a scale $Q_0^2$ large enough for PQCD to be applicable, one can calculate the distributions at any higher $Q^2$.

However, this starting $x$-shape, which depends on non-perturbative
QCD dynamics of the bound state hadron, has not yet been successfully
derived from first principles. Instead they are obtained by fitting
parameterizations to data, in particular structure function
measurements in deep inelastic lepton-nucleon scattering (DIS), \eg\
the CTEQ~\cite{CTEQ} and MRST~\cite{MRST} parameterizations.

Here we present further developments of our phenomenological model
\cite{Edin:1998dz} used to derive the parton distributions from simple
assumptions regarding the non-perturbative properties of the
hadron. The basic idea is to define the valence parton momentum
distributions in the hadron rest frame where we assume that they are
described by spherically symmetric Gaussians. The typical width of these distributions is a few hundred MeV from the Heisenberg uncertainty relation applied to the hadron size. Sea partons are described through quantum
fluctuations of the nucleon into baryon-meson pairs, having the same
quantum numbers as the original nucleon, and the sea parton is
identified with a valence parton in the fluctuation meson. Through
this mechanism, asymmetries in the non-perturbative sea parton
distributions naturally arise, such as the difference between the
$\bar u$ and $\bar d$ distributions and an asymmetry in the momentum
distribution of $s$ compared to $\bar s$.

The explanation of asymmetries in the nucleon sea by fluctuations of
the nucleon into baryon-meson pairs also occurs in ``meson cloud''
models as reviewed in \cite{Kumano-Vogt}. Although the meson cloud
models have an elaborate theoretical formalism (see \eg\
\cite{Holtmann:1996be}) one is forced to introduce phenomenological
parts in order to obtain numerical results that can describe
data. This concerns the choice of form factor in the nucleon to
baryon-meson ($N\to BM$) splitting function and the parton
distributions in both mesons and baryons. Our model invoke similar
hadronic fluctuations as a basic quantum phenomenon, but differs from
meson cloud models in important aspects. We use Gaussian momentum
distributions both for hadrons in the fluctuations and for the partons
in the hadrons, and the hadronic fluctuations are only considered at the
low starting scale $Q^2_0$, where the starting parton densities are
defined and then evolved using standard perturbative QCD
evolution. Thus, we have physically motivated parton momentum
distributions in the nucleon as well as in the fluctuation hadrons,
while meson cloud models usually use parameterizations. Our model
results in a complete set of parton density functions, which we
compare with different experimental data sets giving insights into
different aspects of the non-perturbative nucleon structure.

In section 2 we develop the details of the model and in section 3 we
compare to the relevant experimental data as well as to conventional parameterizations of parton densities. Finally, section 4 gives a concluding discussion. 

%%%%%%%%%%%%%%%%%%%
\section{The model}
%%%%%%%%%%%%%%%%%%%

%%%%%%%%%%%%%%%%%%%%%%%%%%%%%%%%%%
\subsection{Valence distributions}

\begin{figure}[th]
\begin{center}
\includegraphics*[width=4cm]{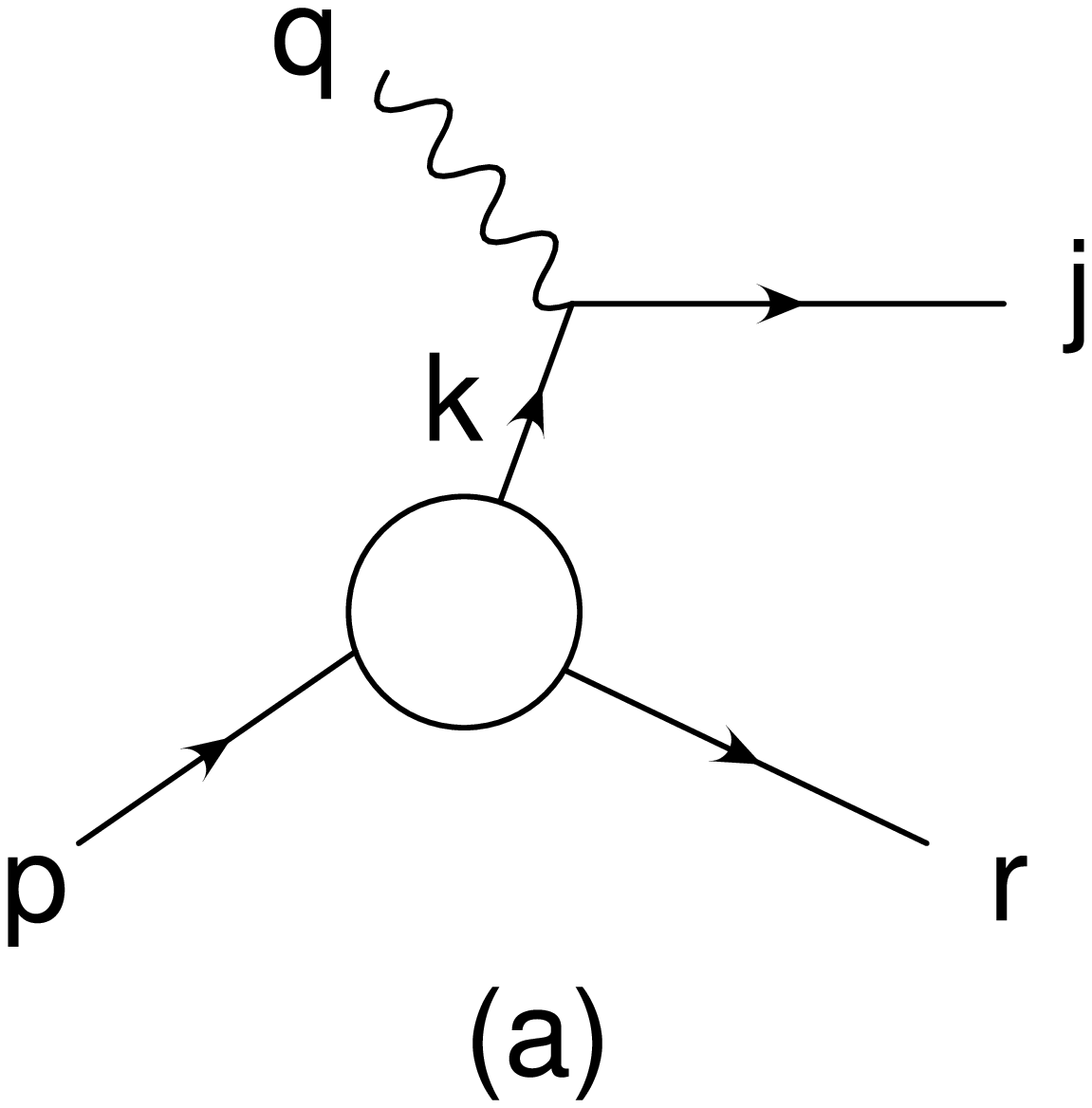}
\includegraphics*[width=4cm]{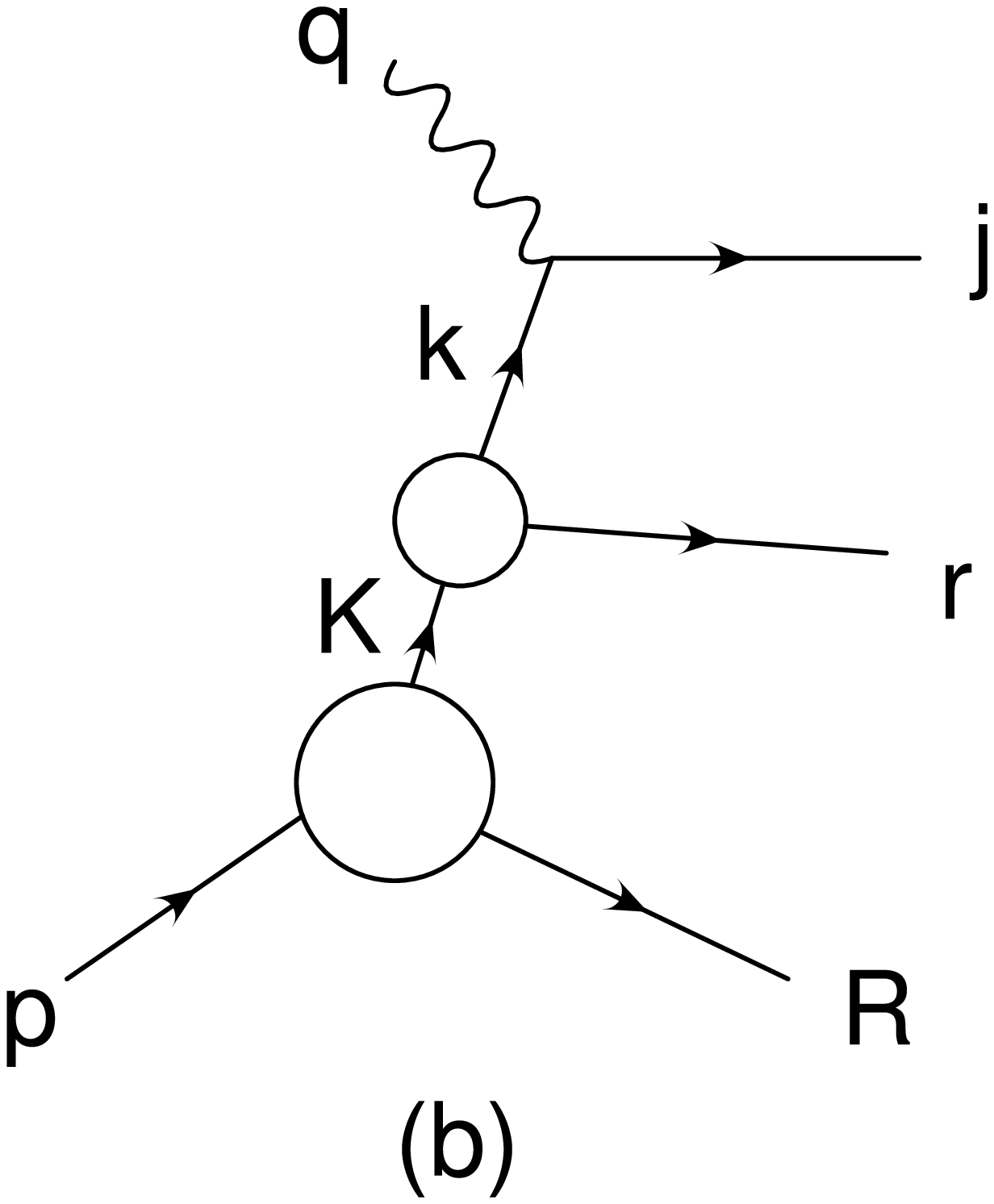}
\vspace*{-15mm}\caption{\label{fig:diagrams} Probing a valence parton in the proton
and a sea parton in a hadronic fluctuation (letters are
four-momenta).}
\end{center}
\end{figure}

This work is an extension of the previously presented physical model
\cite{Edin:1998dz}, giving the momentum distributions of partons in
the nucleon, as illustrated in Fig.~\ref{fig:diagrams}. The model
gives the $x$-shape of the parton distributions of a hadron of mass
$M$ at a momentum transfer scale $Q_0\approx 1\GeV$, \ie\
$xq(x,Q_0^2)$ and $xg(x,Q_0^2)$, which provide an effective
description of the non-perturbative dynamics of the bound state
nucleon. Our approach is not intended to provide the full wave
function for the hadron, but gives the four-momentum $k$ of a single
probed parton, whereas all other partons are treated collectively as a
single remnant with four-momentum $r$ (see
Fig.~\ref{fig:diagrams}). In the nucleon rest frame, there is no
preferred direction and hence the parton momentum distribution is
spherically symmetric. The shape of the momentum distribution for a
parton of type $i$ and mass $m_i$ is then taken as a Gaussian
\begin{equation} \label{eq:gaussian}
f_i(k) = N(\sigma_i,m_i) \exp\left(-\textstyle\frac{(k_0-m_i)^2+k_x^2+k_y^2+k_z^2}{2\sigma_i^2}\right)\;,
\end{equation}
where $N$ is the normalization. The Gaussian form is not only a
reasonable first guess, but can be motivated as the collective result
of the many small momentum transfers affecting the parton through the
non-perturbative bound state interactions which cannot be calculated
properly. The width $\sigma_i$ of the distribution should be of order
hundred MeV from the Heisenberg uncertainty relation applied to the
nucleon size, \ie\ $\sigma_i=1/d_N$. This Fermi motion inside the
nucleon provides the `primordial transverse momentum', which has been
extracted from deep inelastic scattering data and found to be well
described by a Gaussian distribution of a few hundred MeV width
\cite{kt} giving phenomenological support for this description.

The energy component does not have the same simple connection to the
Heisenberg uncertainty relation. To keep the model simple and reduce
the number of parameters, we assume a Gaussian
distribution around the parton mass with the same width as the
momentum fluctuations, such that partons can be off-shell
at a soft scale of the binding interactions. This means a parton
fluctuation life-time corresponding to the nucleon radius.

The momentum fraction $x$ of the parton is then defined as the
light-cone fraction $x=k_+/p_+$. Here, four-momenta are expressed as
$p=(p_+,p_-,\vec{p}_\perp)$ where the `plus' and `minus' components
are $p_\pm=E\pm p_z$ and the $z$-axis defined by the probe. The
fraction $x$ is then invariant under boosts along the $z$-axis and
equivalent to the conventional momentum fraction $x=k_z/p_z$ in a
frame where $p_z$ is large (`infinite momentum' frame).

In order to obtain a kinematically allowed final state, one must
impose the following constraints. The scattered parton must be
on-shell or have a time-like virtuality (causing final state QCD
radiation), and an invariant mass less than the invariant mass of the
hadronic system. Furthermore,
the hadron remnant $r$ is obtained from energy-momentum conservation
and must be time-like in order to project it on a final hadronic state through hadronization. Referring to Fig.~\ref{fig:diagrams}a for the definitions of the momenta, one thus obtains:
\begin{subequations}
\label{eq:limits}
\begin{equation}
m_i^2 \le j^2 = (k+q)^2 < W^2 = (p+q)^2
\end{equation}
\begin{equation}
r^2=(p-k)^2>0
\end{equation}
\end{subequations}
An important consequence of these constraints is to ensure that $0<x<1$ and $f(x)\to 0$ as $x\to 1$.

The parton distributions are obtained by integrating
eq.~(\ref{eq:gaussian}) with these conditions. Using a Monte Carlo
method this can be achieved numerically without approximations. For
the simple case of the valence distributions it is possible to derive
an analytical expression as well, see sec.~\ref{sec:analytical} below.
The normalization of the valence distributions is provided by the sum
rules
\begin{equation} \label{eq:flavoursumrule}
\int_0^1 dx\; u_v(x) = 2 \;\;\; {\rm and} \;\;\; \int_0^1 dx\; d_v(x) = 1 
\end{equation}
to get the correct quantum numbers of the proton (and similarly for
other hadrons). The normalization of the gluon distribution is chosen
so as to saturate the momentum sum rule
\begin{equation}\label{eq:mom-sumrule}
\sum_i \int_0^1 dx \; xf_i(x) = 1\;.
\end{equation}
where the sum is over the valence quarks and gluons.

%%%%%%%%%%%%%%%%%%%%%%%%%%%%%%%%%%%%%%%%%%%%%%%%%%%%%%%%%%%%%%%%%%%%%%%%%%
\subsection{Analytical expressions}
\label{sec:analytical}

By integrating $f(k)$ in eq.~\eqref{eq:gaussian} over $k_\perp^2$ and
$k_-$ of the parton between the limits given by \eqref{eq:limits}, we
arrive at the following analytical expression for the bare hadron
valence distributions:

\begin{eqnarray}
\label{eq:analytic}
f_i(x)&=&N(\tilde\sigma_i,x_i)\left\{
\left[1+\erf\left(\frac{1-x_i}{2\tilde\sigma_i}\right)\right]
\exp\left[-\frac{(x-x_i)^2}{4\tilde\sigma_i^2}\right]-\right.\nonumber\\
&&\left.\left[1+\erf\left(\frac{x-x_i}{2\tilde\sigma_i}\right)\right]
\exp\left[-\frac{(1-x_i)^2}{4\tilde\sigma_i^2}\right]\right\},
\end{eqnarray}
where $\tilde\sigma_i\equiv\sigma_i/M$, $x_i\equiv m_i/M$ and
$\erf(x)=\frac{2}{\sqrt{\pi}}\int_0^xdt\,e^{-t^2}$ (the error
function). The normalization $N(\tilde\sigma_i,x_i)$ is given by
\eqref{eq:flavoursumrule}. This analytical expression can be
simplified when $\sigma_i$ and $m_i$ are sufficiently small (\eg\
$\sigma_i,\,m_i\lesssim M/4$) to
\begin{equation}
\label{eq:approx}
f_i(x)=N(\tilde\sigma_i,x_i)
\left[1-\exp\left(\frac{(1-x_i)(1-x)}{2\tilde\sigma_i^2}\right)\right]
\exp\left(-\frac{(x-x_i)^2}{4\tilde\sigma_i^2}\right)
\end{equation}

\begin{figure}[htbp]
\begin{center}
\includegraphics*[width=8cm]{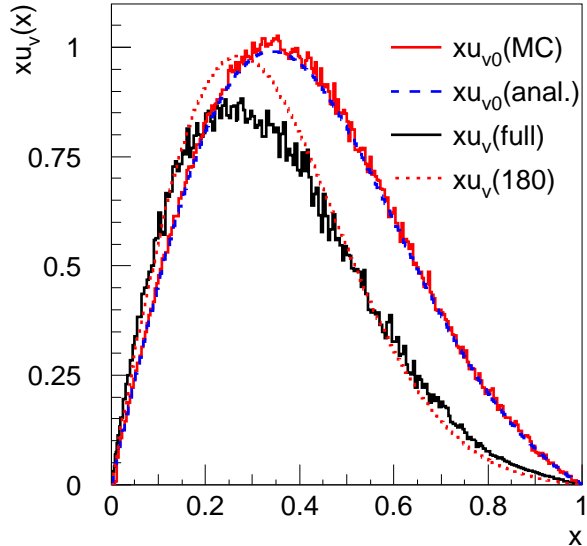}
\caption{\label{fig:analyt} Valence $u$-quark distribution of the
proton obtained from the model: the ``bare'' proton distribution
$xu_{v0}(x)$ from Monte Carlo simulation compared to the analytical
form eq.\ \eqref{eq:analytic}, as well as the ``full''
$u_v$-distribution including simulation of partons in baryons in
hadronic fluctuations $|p\rangle \to |BM\rangle$ as described in
sec.~\ref{sec:sea} below. The dotted line is
the valence distribution of \cite{Edin:1998dz}, having a lower width
$\sigma_u=180\MeV$ but no effect of baryons in fluctuations. (All curves
show the momentum density, whereas the normalization is in the number
density according to eq.~\eqref{eq:flavoursumrule}.)}
\end{center}
\end{figure}

The ``bare'' proton valence $u$-quark momentum distribution $xu_{0v}(x)$
given by eq.~\eqref{eq:analytic} is shown by the full curve in Fig.~\ref{fig:analyt}. The approximation in eq.~\eqref{eq:approx} gives an indistinguishable result and is therefore not shown explicitly. Fig.~\ref{fig:analyt} also demonstrates a very close agreement with the Monte Carlo simulation result of the model (histogram). 

This behavior of the valence distributions is modified by the
contributions from the baryons in the hadronic fluctuations described
below. At $x \gtrsim 0.7$ the shape of the valence
distributions are still given exclusively by the original hadron, \ie\ by
eq.~\eqref{eq:analytic}, but at lower $x$ the valence distributions are significantly modified. The shift of the distribution to lower $x$ is clearly seen in Fig.~\ref{fig:analyt} as the difference between the ``bare'' $xu_{0v}(x)$ distribution and the ``full'' $xu_{v}(x)$ valence $u$-quark momentum distributions. A similar end result can also be obtained without including this baryon fluctuation effect, as in \cite{Edin:1998dz}, by a lower Gaussian width when fitting the model to $F_2$ data (discussed below in section \ref{sec:f2}). 

One can note that the asymptotic behavior for very large $x\gtrsim 0.9$ is $(1-x)^1$ according to eq.~\eqref{eq:analytic}. For $0.3\lesssim x\lesssim 0.9$ however, the $u$ quark distribution is well described by a form $u(x)=8(1-x)^{2.8}$, in good agreement with the $(1-x)^3$-dependence predicted by spectator counting rules \cite{counting-rules}.

%%%%%%%%%%%%%%%%%%%%%%%%%%%%%%%%%%%%%%%%%%%%%%%%%%%%%%%%%%%%%%%%%%%%%%%%%%
\subsection{Hadronic fluctuations and sea distributions}
\label{sec:sea}

Sea partons arise from the non-perturbative dynamics of the bound
state hadron, for which it should be appropriate to use a hadronic quantum
mechanical basis. Therefore we consider hadronic fluctuations of the proton
\begin{equation} \label{eq:hadronfluctuations}
|p\rangle = \alpha_0|p_0\rangle + \alpha_{p\pi}|p\pi^0\rangle + \alpha_{n\pi}|n\pi^+\rangle + \ldots + \alpha_{\Lambda K}|\Lambda K^+\rangle + \ldots 
\end{equation}
where the different states are assumed to be orthonormal and the
normalization constants are real and fulfill
$\alpha_0^2+\sum_{BM}\alpha_{BM}^2=1$.

Probing a parton $i$ in a hadron $H$ of such a fluctuation
(Fig.~\ref{fig:diagrams}b) gives a sea parton with light-cone fraction
$x=x_H\, x_p$ of the target proton, \ie\ the sea distributions are
obtained from a convolution of the momentum $K$ of the hadron and the
momentum $k$ of the parton in that hadron. We assume that the momentum
$\vec{K}$ of the probed hadron is given by a similar Gaussian as in
eq.~\eqref{eq:gaussian}, with a separate width parameter $\sigma_H$
(for simplicity and definiteness we make this parameter common for all
fluctuations). The momentum $\vec{K}^\prime$ of the other hadron in
the fluctuation is then fixed by momentum conservation in the rest
frame of the original nucleon. We let both hadrons be on-shell, which
fixes their energies. This implies that energy is not conserved at
this intermediate stage, but is of course restored for the observable
final state. With the hadron four-vectors specified, one obtains the
light-cone fraction $x_H=K_+/(K+K^\prime)_+$. A few possible
modifications of the details of the model are discussed in the next
section and the possible use of an effective meson mass instead of the
physical pion mass is discussed in sec.~\ref{sec:dubar}.

The above model for valence distributions is then applied to the
fluctuation hadron $H$ to get the parton momentum and light-cone fraction
$x_p=k_+/K_+$ in $H$. The flavor sum rules in eq.~(\ref{eq:flavoursumrule})
must, of course, be modified to apply for $H$. The kinematical constraints of eq.\ (\ref{eq:limits}) are in this case modified to 
\begin{subequations}
\label{eq:newlimits}
\begin{equation}
m_i^2\le j^2 < W_H^2=(K+q)^2 
\end{equation}
\begin{equation}
r^2>0 \;\; \mathrm{and} \;\;(r+R)^2=(p-k)^2>0
\end{equation}
\end{subequations}
such that the scattered parton and the remnants (\cf\ Fig.~\ref{fig:diagrams}b) are on-shell or have positive virtualities within physically allowed limits and hence can give a proper final hadronic state. 

A Monte Carlo method is
used to simulate this two-step process by choosing $K$ and $k$, impose
the constraints and obtain the momentum fraction $x$. By iterating the
procedure the additional quark and gluon distributions due to the
fluctuations are generated. Note that the flavor number density
$\int_0^1dx\,f_i(x)$ is not affected by the convolution with the
fluctuation momentum. This means that the normalization criteria
\eqref{eq:flavoursumrule} (modified for the hadron in question)
automatically ensure flavor conservation, since the meson and the
baryon in the fluctuation are multiplied by the same normalization
constant $\alpha_{BM}^2$.

Since the $x$ distribution of the fluctuation hadrons is given by
$x_H=K_+/(K+K^\prime)_+\approx M_H/(M_B+M_M)$,
the low-$x$ sea is mainly given by scattering off the meson
in the baryon-meson fluctuation. The scattering
off the baryon in the baryon-meson fluctuation gives a contribution
resembling the valence distribution, but at lower $x$ than the bare
hadron distribution. This means that the analytical valence
distribution of the bare hadron derived above is modified for
$x\lesssim 0.7$ with the inclusion of hadronic fluctuations, as shown in Fig.~\ref{fig:analyt}.

The normalization of the sea distributions is given by the amplitude
coefficients $\alpha_{BM}^2$. These partly depend on Clebsch-Gordan
coefficients and should also include a suppression for larger masses. However, the full dependence is due to non-perturbative
dynamics that cannot be calculated from first principles in QCD,
including the possibility of mixing of different mesons/baryons with
the same quark content. We therefore take the normalizations
$\alpha_{BM}^2$ as free parameters.

%%%%%%%%%%%%%%%%%%%%%%%%%%%%%%%%%%%%%%%%%%%%%%%%%%%%%%%%%%%%%%%%%%%%%%
\subsubsection{Alternatives in the definitions of fluctuation momenta}
\label{sec:alternatives}

In the details of the model described above we have made some choices which could be done differently:

\begin{enumerate}
\item In the definition of the fluctuation momentum, the fluctuation hadrons
were made on-shell. Another possibility would be to let the energy of
the probed hadron fluctuate around the hadron mass as in
eq.~\eqref{eq:gaussian}, allowing the fluctuation hadrons to be
off-shell on a soft scale. The differences in the resulting sea
distributions are however small and can be mimicked by changing
parameter values.

\item \label{item:upper} We defined the upper bound on the
scattered parton virtuality as $j^2<W_H^2=(K+q)^2$. This choice keeps the
parton distributions in the fluctuation hadrons close to unaltered
with respect to the distributions in the corresponding free
hadrons. However, this choice allows the struck parton to have
essentially all the proton momentum, which might seem unnatural. An
alternative approach would be to put the upper bound on $j^2$ to $(x_H
P+q)^2\simeq x_HW^2$, such that it is restricted by the longitudinal
momentum fraction taken by the fluctuation hadron. This results in a
softer meson momentum spectrum, because the distributions in the
fluctuation meson is distorted. For completeness, we will consider
both alternatives in our predictions for the $s-\bar s$ asymmetry
(sec.~\ref{sec:ssbar}).

\item The limits on the remnant momenta was put to $r^2>0$ (to ensure
that $f(x)\to 0$ as $x\to 1$) and $(r+R)^2=(p-k)^2>0$. This latter
choice is motivated by the fact that the fluctuation hadrons are close in space, such that an exchange of momentum on a soft
scale should be possible and therefore the condition is on the whole remnant system. Another possibility would be to force $r$
and $R$ separately to be time-like. This is obviously not possible if
we scatter on a baryon with mass larger than the original hadron mass
and make the fluctuation hadrons on-shell, since $R^2>0$ can then
never be satisfied. In the off-shell case discussed under point 1 this is possible, and gives slightly harder meson spectra for low meson
masses, whereas for meson masses $\gtrsim 300\MeV$ the difference is
negligible.
\end{enumerate}

%%%%%%%%%%%%%%%%%%%%%%%%%%%%%%%%%%%%%%%%%%%%
\subsection{GVDM contributions at low $Q^2$}

\begin{figure}[th]
\begin{center}
\includegraphics*[width=8cm]{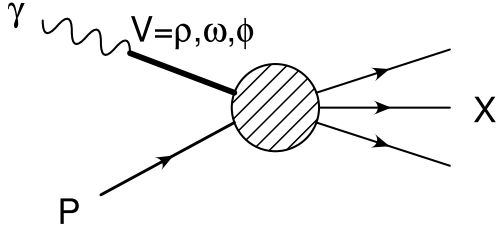}
\vspace*{-5cm}\caption{\label{fig:vdm} Schematic view of the photon 
fluctuating to a vector meson before interacting with the proton.}
\end{center}
\end{figure}

In the same spirit as the hadronic fluctuations of the proton
described above in eq.~(\ref{eq:hadronfluctuations}), one should also
note that a photon may appear as a vector meson, such that the quantum
state should be expressed as
\begin{equation}
\label{eq:photon-fluctuation}
|\gamma\rangle = C_0|\gamma_0\rangle + \sum_V \frac{e}{f_V}|V\rangle + \int_{m_0}dm (\cdots)
\end{equation}

In the original vector meson dominance model (VDM) only fluctuations to
vector mesons were considered \cite{VDM}, but in generalized models
(GVDM), also a contribution from a continuum of higher-mass states is
included \cite{GVDM}, represented in eq.~\eqref{eq:photon-fluctuation}
by an integral over masses.

The GVDM interpretation of the photon-proton interaction is
illustrated in fig.~\ref{fig:vdm}. The corresponding cross section is
given by a convolution of the photon-to-meson fluctuation probability
with the meson propagator and the meson-proton cross section
\cite{GVDM}, and the continuum contribution is included using a
phenomenologically chosen spectral weight function. As shown in
\cite{Alwall:2004wk}, the resulting expression for the proton
structure function,
\begin{eqnarray}\label{eq:F2-GVDM}
F_2(x,Q^2) & = & \frac{(1-x)Q^2}{4\pi^2\alpha} 
      \left\{  \sum_{V=\rho, \omega, \phi} r_V \left(\frac{m_V^2}{Q^2 + m_V^2}\right)^2
	\left(1 + \xi_V\frac{Q^2}{m_V^2}\right) \right.
\nonumber \\
 & & \left. +\; r_C\left[ (1-\xi_C)\frac{m_0^2}{Q^2 + m_0^2} + 
        \xi_C \frac{m_0^2}{Q^2}\ln{(1 + \frac{Q^2}{m_0^2})}
	\right] \right\}
	A_\gamma \frac{Q^{2\epsilon}}{x^\epsilon}\;,
\end{eqnarray}
describes data very well for $Q^2 \lesssim 0.7 \GeV^2$.

At higher $Q^2$ this GVDM contribution should be phased out in order
to conform to the conventional description in terms of parton density
functions. One way of doing this \cite{Alwall:2004wk} is by
introducing a phenomenological form factor $(Q^2_C/Q^2)^a$ on the GVDM
contribution for $Q^2>Q^2_C$, which gives a good description of HERA
$F_2$ data at these intermediate $Q^2 \lesssim 4\GeV^2$. The GVDM
component was found to be negligible for $Q^2\gtrsim 4\GeV^2$.
 
The details of the phasing out of the GVDM component is not known and would require additional assumptions and parameters in the model. Fortunately, GVDM (giving symmetric quark-antiquark contributions) can be neglected in this study of the asymmetries in the proton, since the relevant data used in the following is at $Q^2\gtrsim 4\GeV^2$ where the GVDM component was found to be negligibly small.

%%%%%%%%%%%%%%%%%%%%%%%%%%%%%%%%%%%%%%%%%%%%%%%
\section{Results and comparisons to experiment}
%%%%%%%%%%%%%%%%%%%%%%%%%%%%%%%%%%%%%%%%%%%%%%%

With this simple model, based only on Gaussian momentum fluctuations
and hadronic quantum fluctuations together with kinematical constraints, we are able to describe several 
different aspects of the parton distribution functions of the
nucleon. Perhaps phenomenologically most interesting, we describe the
asymmetries between $\bar u$ and $\bar d$ in the proton quark
sea. Also, we get an asymmetry in the nucleon strange sea, such that
the $s$ quark has a harder momentum distribution than the $\bar s$
antiquark, which is of interest \cite{Alwall:2004rd} in relation to
the \nutev\ anomaly \cite{nutev}.

Since we have most data on the proton, the parton widths are fitted
for this special case of a hadron. For other hadrons, we assume for
simplicity that the widths of the gluon distribution is the same as
that for the proton. For valence quarks, we assume that the $d$ quark
width is the typical width for a quark with quark number $\int
q(x)dx=1$ in a hadron, while the $u$ quark width is typical for a
quark with number $\int q(x)dx=2$. This means that in all mesons and
also \eg\ the $\Lambda$ baryon, all valence quark widths are given by
the proton $d$ quark width. This can be motivated by the hadrons
having essentially the same spacial extent, making the only relevant
difference the possible effects of the Pauli exclusion principle. Such
'Pauli blocking' (see \cite{Kumano-Vogt} and references therein) would reduce the effective space available for the $u$ quarks in the proton, resulting in a larger momentum fluctuation width. Therefore, for the neutron we simply make an isospin transformation $d\leftrightarrow u$ of the proton distribution parameters.

\begin{figure}
\begin{center}
\includegraphics*[width=8cm]{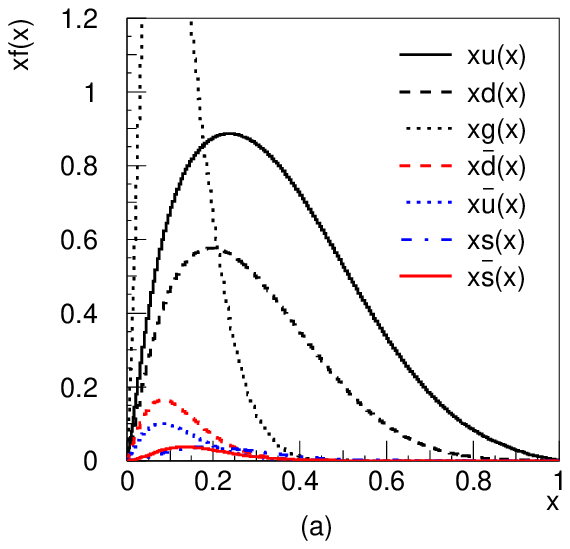}
\includegraphics*[width=8cm]{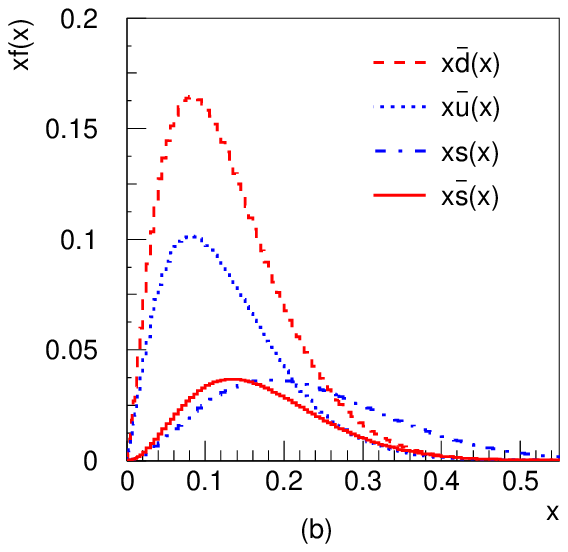}
\caption{\label{fig:pdfs} Parton distributions from our model at the
starting scale $Q_0^2$: (a) all partons distributions (b) the sea parton
distributions (note the different scales).}
\end{center}
\end{figure}

The model provides valence and sea parton $x$-distributions as shown
in Fig.~\ref{fig:pdfs}. These apply at the low starting scale $Q_0^2$,
while the parton distributions and the proton structure function
$F_2(x,Q^2)$ at higher $Q^2$ are obtained by applying standard QCD evolution as implemented in the QCDNUM16 code \cite{Botje}. For the results below we have used the conventional $\overline{MS}$ scheme and the evolution equations in next-to-leading order to have the highest available theoretical accuracy. Using the option of only leading order equations, the values of all model parameters are unaffected except $Q_0$ and $\sigma_g$ as discussed in section~\ref{sec:cteq}. With leading order evolution, however, the quality of the fit to data becomes less good.  

Since different model parameters are sensitive to different data, we
have used data from several different experiments: 
\begin{itemize}
\item Fixed-target $F_2$ data \cite{NMC,BCDMS} to fix large-$x$
(valence) distributions
\item HERA $F_2$ data \cite{H1} for the gluon distribution width
and the starting scale $Q_0^2$
\item $W^\pm$ charge asymmetry data \cite{w-asym} as a cross-check on
the ratio of Gaussian widths for the $u$ and $d$ valence quark
distributions
\item $\bar d / \bar u$-asymmetry data \cite{dubar} for the
normalizations of the $|p\pi^0\rangle$ and $|n\pi^+\rangle$
fluctuations
\item Strange sea data \cite{ssbar} to fix the normalization of the 
fluctuations including strange quarks. 
\end{itemize}
It should be noted, however, that most of the model parameters
influence several observables at least to some degree, making the task
of a total fit quite involved.

It is interesting to note that this simple model can describe such a
wealth of different data with just one or two parameters per data set,
as will be seen in detail below. The best-fit parameters are:
\begin{equation}
\label{eq:params}
\begin{array}{ccc}
\sigma_u=230\MeV & \sigma_d=170\MeV & \sigma_g=77\MeV\\
Q_0=0.75\GeV & \sigma_H=100\MeV\\
\alpha_\mathrm{p\pi}^2=0.45 & \alpha_{n\pi}^2=0.14 & 
\alpha_{\Lambda K}^2=0.05
\end{array}
\end{equation}

The normalization $\alpha_{BM}^2$ for the proton fluctuating to a
baryon $B$ and a meson $M$ is defined as the fraction of the proton
momentum taken by all partons $i$ in the fluctuation $BM$,
\begin{equation} 
\alpha_{BM}^2=\int_0^1 dx\,\left[\, \sum_{i\in B}xf_{i/B}(x)+  
\sum_{i\in M} xf_{i/M}(x)\right]
\end{equation}

%%%%%%%%%%%%%%%%%%%%%%%%%%%%%%%%%
\subsection{Inclusive $F_2$ data}
\label{sec:f2}

\begin{figure}
\begin{center}
\includegraphics*[width=14cm]{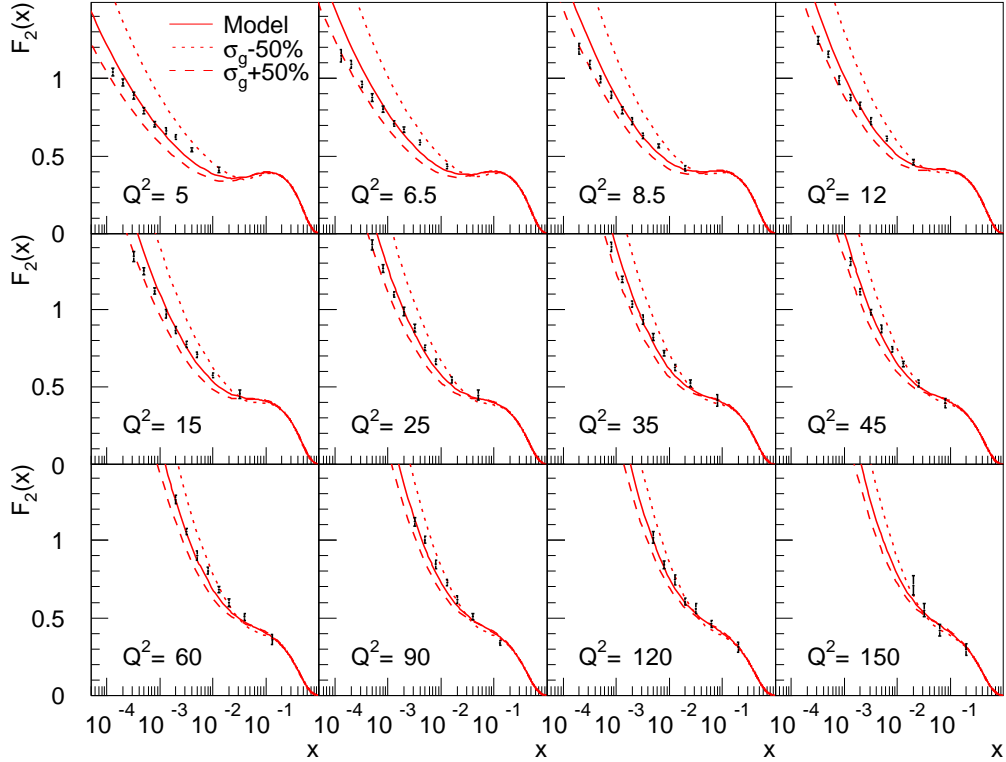}
\caption{\label{fig:f2} The proton structure function $F_2(x,Q^2)$;  
H1 data \cite{H1} compared to our model with its sensitivity to a $\pm50\%$ variation of the width parameter $\sigma_g$ of the gluon distribution.}
\end{center}
\end{figure}
\begin{figure}
\begin{center}
\includegraphics*[width=14cm]{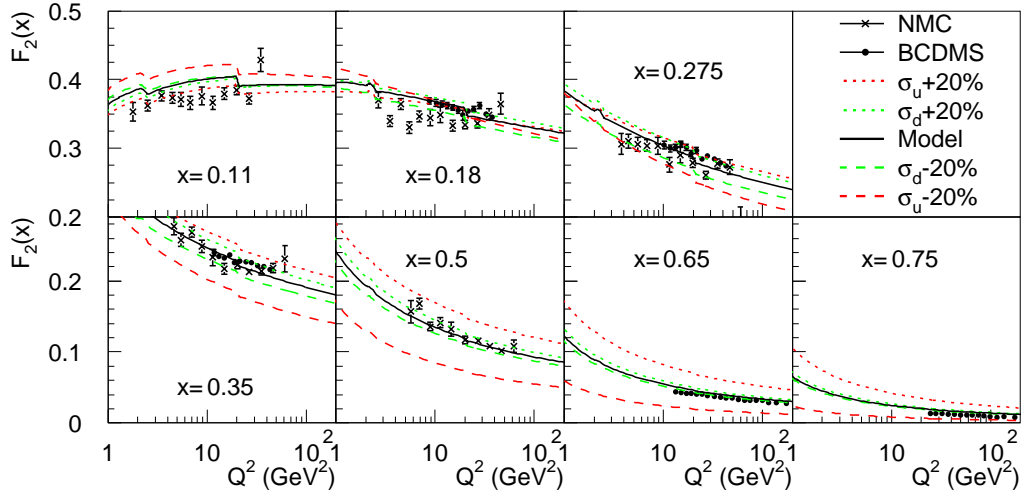}
\caption{\label{fig:xbinned} The proton structure function $F_2(x,Q^2)$ for large $x$ values; NMC and BCDMS data \cite{NMC,BCDMS} compared to our model,
also showing the results of $\pm 20\%$ variations of the width parameters $\sigma_u$ and $\sigma_d$ for the $u$ and $d$ valence distributions.}
\end{center}
\end{figure}

The model was originally \cite{Edin:1998dz} used only to fit inclusive
data, using a somewhat different definition for the hadronic
fluctuations, and was shown to give a good fit. Now we are mainly
interested in the model's description of asymmetries in the proton,
and hence the $F_2$ data is used mainly to fix some model parameters
which are not well restricted by asymmetry data, namely the Gaussian
width of the gluon distribution, $\sigma_g$, and the starting scale
$Q_0$. However, in view of the simplicity of the model, we still get a
nice description of $F_2$ data from H1 \cite{H1}, see
Fig.~\ref{fig:f2}. To show the sensitivity to the value of $\sigma_g$,
we also show the result of varying this parameter with 
$\pm50\% 
$. 
The surprisingly low value for $\sigma_g$ obtained in the fit will be
discussed in sec.~\ref{sec:cteq}.

Since this data is restricted to relatively low $x$ values, we also
compare to fixed-target NMC and BCDMS data \cite{NMC,BCDMS}, which
constrains the Gaussian widths of the $u$ and $d$ quark distributions,
$\sigma_u$ and $\sigma_d$. This is shown in Fig.~\ref{fig:xbinned}
which also shows the sensitivity of the fit to changing these
parameters with 
$\pm20\% 
$. 
As seen in the figure, the $u$ width is quite well constrained by the large-$x$ data, while the $d$ width is less well constrained, and the data might account for a $\sigma_d$ which is up to $20\%
$ 
smaller. This is certainly true for the NMC data, which actually seems to prefer a smaller $d$ width. However, the full curve describes the best-fit value due to the very small error bars on the BCDMS data, so we keep this as our main result in the following.

%%%%%%%%%%%%%%%%%%%%%%%%%%%%%%%%%%%%%%%%%
\subsection{The $W^\pm$ forward-backward asymmetry}

\begin{figure}
\begin{center}
\includegraphics*[width=8cm]{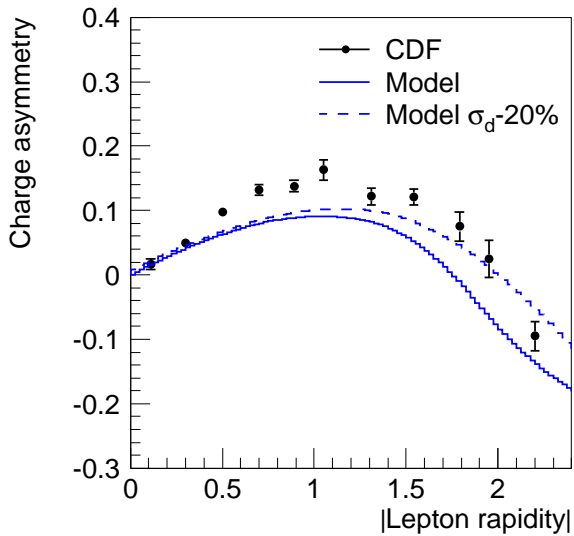}
\vspace*{-0.7cm}
\caption{\label{fig:w-asym} The charge asymmetry versus rapidity, eq.~(\ref{eq:charge-asym}), for leptons from $W^\pm$-decays 
in $p\bar p$ collisions at the Tevatron \cite{w-asym} (with positive
and negative lepton rapidity combined) compared to \pythia\ 
simulations using parton densities from our model, with best-fit
parameters and a $20\%$ reduced width of the valence $d$ quark distribution.}
\end{center}
\end{figure}

The forward-backward charged lepton asymmetry from $W^\pm$ decays in
$p\bar p$ collisions at the Tevatron provides information on
the large-$x$ distributions of $d$ and $u$ quarks. This is because the
$W$ bosons are produced mainly through the processes $u \bar d\to W^+$
and $d\bar u\to W^-$. If the $u$ distribution is harder than the $d$
distribution, we will get more $W^+$ than $W^-$ in the direction of
the proton beam, and vice versa in the direction of the $\bar p$-beam,
which is precisely what is observed. The data is for $Q^2\gtrsim
M_W^2$ and $0.006<x<0.34$, corresponding via PQCD evolution to
$0.01\lesssim x\lesssim 0.6$ at the starting scale $Q_0^2\sim
1\GeV^2$. We can therefore use the data on the charge asymmetry versus
rapidity, expressed as
\begin{equation} \label{eq:charge-asym}
A(y_l)=\frac{d\sigma^+/dy_l \; - \; d\sigma^-/dy_l}{d\sigma^+/dy_l \; + \; d\sigma^-/dy_l}
\end{equation}
to get further information on the Gaussian width of the $d$ quark
distribution. The result, using our PQCD-evolved parton distributions
in \pythia\ to simulate $W^\pm$ production via the processes $q\bar
q'\to W^\pm$, $q\bar q'\to gW^\pm$ and $qg\to W^\pm q'$, and subsequent
decay $W^\pm\to l^\pm\nu_l$, is shown in Fig.~\ref{fig:w-asym}. It is
clear that the model reproduces the salient features of the data, in
particular when the $d$ width is reduced as preferred by the NMC data
discussed above.

The difference between the widths of the $u$ and $d$ quarks
($\sigma_u=230\MeV$ and $\sigma_d=135-170\MeV$) 
can perhaps be attributed to the Pauli blocking effect on the $u$ quarks as discussed above. Large differences in the intrinsic transverse
momentum between the $u$ and $d$ quarks are indeed indicated by data from HERMES \cite{Jgoun:2001ck}. Also the standard parameterizations of parton densities \cite{CTEQ,MRST} result in harder spectra for the $u$ than for the $d$ valence distribution.

%%%%%%%%%%%%%%%%%%%%%%%%%%%%%%%%%%%%%%%%%%%%%%
\subsection{The $\bar d$ - $\bar u$ asymmetry}

\label{sec:dubar}

Now we come to the important results of the model regarding the sea
quark asymmetries. These result from the fluctuations of the proton to
meson-baryon pairs. The first such asymmetry that we will investigate,
and the one with most experimental data, is the difference between
$\bar u$ and $\bar d$ distributions.

In this model, the $\bar u$ and $\bar d$ sea comes mainly from
fluctuations of the proton into a nucleon and a pion, since these are
the lowest energy fluctuations. This means that we will get an excess
of $\bar d$ over $\bar u$, since $|p\pi^0\rangle$ is symmetric between
$\bar d$ and $\bar u$, while $|n\pi^+\rangle$ contains only $\bar d$
and no $\bar u$. If these were the only available fluctuations we
would get from isospin Clebsch-Gordan coefficients the ratio
$\alpha^2_{p\pi}/\alpha^2_{n\pi}=C^2_{p\pi}/C^2_{n\pi}=1/2$. However,
this is modified by the inclusion of $|\Delta\pi\rangle$-fluctuations
(which have small mass suppression relative to the $|N\pi\rangle$
fluctuations). In the SU(6) scheme, the $|\Delta^{++}\pi^-\rangle$
fluctuation has a larger Clebsch-Gordan coefficient than the
$|\Delta^0\pi^+\rangle$, which increases the relative amount of $\bar
u$ quarks. In our model, we take such higher mass
fluctuations into account by simply fitting the normalization
parameters $\alpha^2_{p\pi}$ and $\alpha^2_{n\pi}$ using experimental
data on the light sea.

The ratio $\bar d/\bar u$ can be extracted by comparing muon pair
production through the Drell-Yan process, $q\bar q\to l^+l^-$, in
$pp$ collisions with that in $pd$ collisions assuming isospin
symmetry. Using parameterizations for the sum $\bar d(x)+\bar u(x)$,
also the difference $\bar d(x)-\bar u(x)$ can be extracted. 

We have used data from the E866 collaboration \cite{dubar} to fit the
parameters relevant for the $\bar d$ and $\bar u$ distributions. These
parameters are the normalizations $\alpha_{p\pi}^2$ and
$\alpha_{n\pi}^2$ and the Gaussian width $\sigma_H$. However, the
sensitivity of the meson parton distributions to $\sigma_H$ is small,
and in particular the position of the peak of the momentum
distributions is quite independent of $\sigma_H$, and depends only on
the mass of the fluctuation hadrons. For small $\sigma_H\lesssim
100\MeV$, also the overall shape of the meson distributions is mainly
determined by the hadron masses. For larger $\sigma_H\gtrsim 100\MeV$,
the parton distributions inside the fluctuation hadrons are altered
due to the kinematical limits in such a way as to keep the convoluted
shape stable. Therefore we take the value of $\sigma_H$ to be
$100\MeV$, which is preferred by the fit.

\begin{figure}
\begin{center}
\includegraphics*[width=14cm]{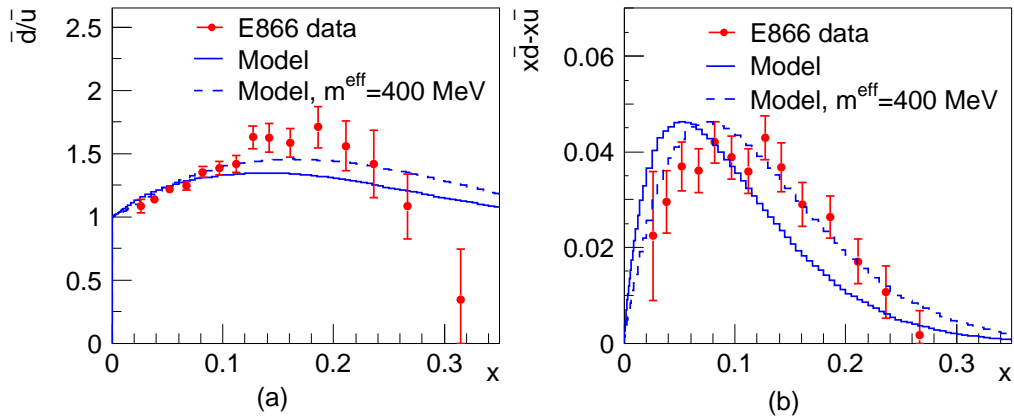}
\caption{\label{fig:dubar} Comparison between our model and data from
the E866/NuSea collaboration \cite{dubar}: (a) $\bar u(x)/\bar d(x)$
(b) $d(x)-u(x)$. The full line shows the results of the model using
the physical pion mass, while the dashed line shows the results
with an effective pions mass $m^\mathrm{eff} = 400\MeV$ as
discussed in the text. The parton distributions are evaluated at
$Q^2=54\GeV^2$.}
\end{center}
\end{figure}

In the distribution of $\bar d(x)-\bar u(x)$ (or, equivalently $x\bar
d(x)-x\bar u(x)$), the effect of symmetric $q\bar q$ pair production
from gluons in the PQCD evolution is canceled and thus the result
depends only on the properties of the non-perturbative distributions at
the starting scale, barring the uncertainties from the extraction of
the data from experiments. Therefore it is quite interesting to
compare the experimental results for this distribution with the model
results. As can be readily seen in Fig.~\ref{fig:dubar}b, the pion
distribution using the physical pion mass $m_\pi=140\MeV$ (full line)
is too soft, peaking at $x\approx 0.05$, while the data peaks around
$x\simeq0.1$. This cannot be cured using a larger $\sigma_H$, as explained
above, but rather indicates that one should use a larger effective
pion mass. Allowing this effective pion mass to be a free parameter in
the fit, the best-fit result is that $m^\mathrm{eff}\approx
400\MeV$, as showed in Fig.~\ref{fig:dubar} (dashed line). Note that
the points driving the fit are the small-$x$ $\bar d/\bar u$ data,
which have the smallest errors.

One can think of several explanations for such a large effective pion
mass. One could be that we need to include a relatively large fraction
of heavier mesons, such as the $\rho$ mesons, in our description. 
This would be in accordance with the finding of a large $|N\rho\rangle$ component in the study \cite{Holtmann:1996be} based on a meson cloud model. 
Another explanation would be that since the pions have
very non-typical hadron masses, the meson-baryon fluctuations involve
more generic meson states, with the quantum numbers defined by the
actual hadron states available, but with masses with some distribution
which can be approximated with a meson mass of $400\MeV$. In any case,
the introduction of a larger pion mass without any other modifications
of the model reduces the quality of the fit to the low-$x$ $F_2$ data
described in sec.~\ref{sec:f2} (the effect is comparable to changing
the value of $\sigma_g$ by less than 
$+50\%
$), which points to the need of introducing more assumptions and parameters to get a perfect description of all data. Since our intention is not to get the best possible fit (as conventional parton density parameterizations), we refrain from doing this without a physics motivation related to our model. 

%%%%%%%%%%%%%%%%%%%%%%%%%%%%%%%%%%%%%%
\subsection{The strange sea asymmetry}

\label{sec:ssbar}

Another consequence of the model is an asymmetry between the strange
and anti-strange quark $x$-distributions in the nucleon, an aspect
that we did a special study of in \cite{Alwall:2004rd} and extend
here. The reason for this asymmetry is that in fluctuations with open
strangeness, the $s$ quark is found in the baryon while the $\bar s$
quark is in the meson. Since the baryon, due to its higher mass, has a
harder $x_H$-spectrum than the meson, we get a harder $x$-spectrum for
the $s$ quark then the $\bar s$. This effect persists in spite of the
fact that the $s$ distribution in the $\Lambda$ peaks at lower $x_p$
than the $\bar{s}$ distribution of the $K^+$, see
\cite{Edin:1998dz} for examples of typical meson and baryon
distributions from the model.

For simplicity
we assume that all fluctuations including strangeness (such as
$|\Lambda K^*\rangle$, $|\Sigma K\rangle$) can be implicitly included
in the $|\Lambda K\rangle$-fluctuation. Fluctuations where the
$s\bar{s}$ pair is part of a meson wave function, thus giving a
symmetric contribution, are neglected since they should be suppressed
($\phi$ due to large mass and $\eta$ due to Clebsch-Gordan
coefficients). The resulting $s$ and $\bar s$ distributions are shown
in Fig.~\ref{fig:pdfs}b.

Note that in this model the criterion that the nucleon should have
zero total strangeness, $\int_0^1 (s(x)-\bar s(x))=0$, is automatically
fulfilled, since all strange fluctuations have the same number of $s$
quarks in the baryon as $\bar s$ quarks in the meson due to the sum
rules \eqref{eq:flavoursumrule} modified for the hadrons in
question. This normalization is not changed when the parton
distributions of the fluctuation are convoluted with the distributions
of the fluctuation hadrons in the nucleon (except for the factor
$\alpha_{BM}$ which is common to the baryon and the meson).

\begin{figure}
\begin{center}
\includegraphics*[width=15cm]{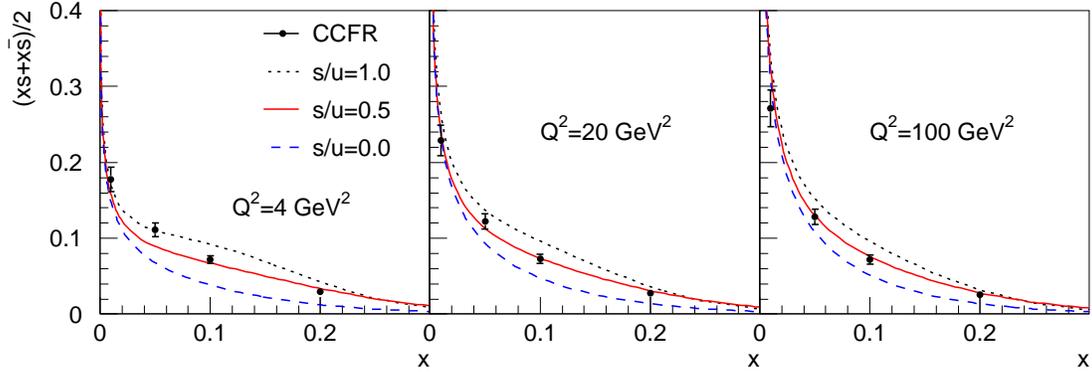}
\caption{\label{fig:sdata} CCFR deep inelastic scattering data 
\cite{ssbar} on the strange sea distribution 
$(xs(x)+x\bar{s}(x))/2$ in the nucleon at different $Q^2$ compared to
our model based on $|\Lambda K\rangle$ fluctuations (with normalization `s/u'$=(s+\bar{s})/(\bar{u}+\bar{d})$ as discussed in the text) at a low scale and evolved to larger $Q^2$ with perturbative QCD that also adds a symmetric perturbative $s\bar{s}$ component.
}
\end{center}
\end{figure}

Since the normalization of the $|\Lambda K+\rangle$ fluctuation cannot
be safely calculated, we take it as a free parameter which we fit to
data from the CCFR collaboration \cite{ssbar} on the averaged strange sea
$(s+\bar s)/2$, as shown in Fig.~\ref{fig:sdata}. 
Here, perturbative QCD evolution to larger $Q^2$ shifts the original $s$ and $\bar{s}$ distributions to smaller $x$ and adds a symmetric sea arising from $g\to s\bar{s}$. The fit gives a
$|\Lambda K+\rangle$ normalization $\alpha_{\Lambda K}^2=0.055$ 
such that
$\int_0^1dx(xs(x)+x\bar{s}(x))/\int_0^1dx(x\bar{u}(x)+x\bar{d}(x))\approx
0.5$, \ie\ the strange sea momentum fraction at $Q_0^2$ is
approximately half of that of a light sea quark, in agreement with the
parton density analyzes in \eg\ \cite{Martin:1998sq} and the meson cloud model study in \cite{Holtmann:1996be}. 
 
This normalization of the non-perturbative strange quark sea means that the coefficient $\alpha_{\Lambda K}^2$ has essentially been scaled down with the fluctuation time $\Delta t \sim 1/\Delta E$ relative to the light sea. This is a smaller suppression than $1/(\Delta E)^2$ that one might have expected based on old-fashioned perturbation theory. On the other hand, this $s\bar{s}$ suppression is of similar magnitude as the ratio $P(s\bar{s})/P(u\bar{u})\approx 1/3$ of the probabilities for quark-antiquark production in phenomenological hadronization models, such as the Lund model \cite{Andersson:ia}. Given that both cases concern non-perturbative $s\bar{s}$ pair production in a color field, this need not be surprising but indicate common features. 

This $|\Lambda K+\rangle$ fluctuation results in an asymmetry $S^- = \int_0^1dx(xs(x)-x\bar s(x))=0.00165$ at $Q^2=20\GeV^2$. 
However, this result is not
altogether independent of the details of the model. The precise
definition for the upper limit of the struck quark virtuality
discussed in sec.~\ref{sec:alternatives}, item~\ref{item:upper}, and
the width of the $d$ distribution (which gives the width of the
valence quark distributions in the $\Lambda$ and $K$) both have
influence on the value. Using both definitions for the limit, and
allowing the $d$ width to vary between the values used in
Fig.~\ref{fig:w-asym}, we get a range of values for the asymmetry
\begin{equation} 
\label{eq:srange}
0.0010\leq S^-\leq 0.0023
\end{equation}

\begin{figure}
\begin{center}
\includegraphics*[width=80mm]{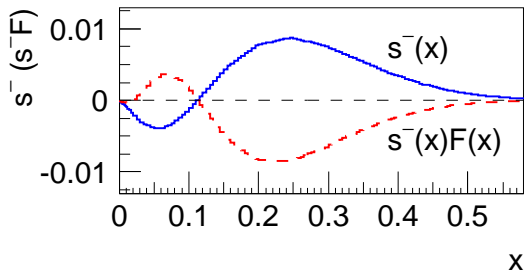}
\vspace*{-5mm}
\caption{\label{fig:sin2theta} The strange sea asymmetry $s^-(x) = 
xs(x)-x\bar{s}(x)$ (at $Q^2=20\GeV^2$) from the model and combined
with the function $F(x)$ accounting for NuTeV's analysis giving
$\Delta \sin^2\theta_W = \int_0^1 dx\, s^-(x) F(x) = -0.0017$.}
\end{center}
\end{figure}

An asymmetry in the momentum distributions of $s$ and $\bar s$ is
especially interesting in view of the \nutev\ anomaly. 
In the \nutev\ experiment \cite{nutev}, the value of
$\sin^2\theta_W$ was extracted from neutral and charged current
cross-sections of neutrinos and anti-neutrinos. The value they find
differs by about $3\sigma$ from the value obtained in Standard Model
fits to data from other experiments: $\sin^2\theta_W^\mathrm{NuTeV} = 0.2277
\pm 0.0016$ while $\sin^2\theta_W^\mathrm{SM}=0.2227 \pm
0.0004$. A number of possible explanations \cite{Davidson:2001ji} for
this discrepancy have been suggested, both in terms of extensions to
the Standard Model and in terms of effects within the Standard
Model. One explanation within the Standard Model is that, since the
neutrinos and anti-neutrinos interact differently with $s$ and $\bar
s$ quarks, an asymmetry $S^-\neq 0$ shifts the \nutev\
result. In order to facilitate the calculation of such a shift,
\nutev\ has published a folding function $F(x)$ \cite{Zeller:2002du},
to account for their analysis and give the shift in the extracted
value of $\sin^2\theta_W$, where the parton densities are to be
calculated at $Q^2=20\GeV^2$. Using this, we get the shift $\Delta
\sin^2\theta_W = \int_0^1 dx\, s^-(x) F(x)= -0.0017$, corresponding to
a reduction of the significance of the anomaly with one standard
deviation. Our results for $s^-(x)=xs(x)-x\bar s(x)$ and $s^-(x)F(x)$
are shown in Fig.~\ref{fig:sin2theta}. The range of values of $S^-$ of
eq.~\eqref{eq:srange} corresponds to
$-0.0024\leq\Delta\sin^2\theta_W\leq-0.00097$ for the shift in
$\sin^2\theta_W^\mathrm{NuTeV}$. This reduces the anomaly to between
$1.6\sigma$ and $2.4\sigma$ away from the Standard Model value.

This important topic has been investigated previously. On the
theoretical side we note that it has recently been shown
\cite{Catani:2004nc} that higher order perturbative effects give a
negative contribution to $S^-$, although significantly smaller than
our positive non-perturbative effect. Within the meson cloud model of
\cite{Cao:2003ny}, based on exponentially suppressed $\Lambda K$
fluctuations, an inconclusive result on the sign of $S^-$ was obtained
and a strange sea that was too small to affect the NuTeV anomaly. The
meson cloud model in \cite{Holtmann:1996be} obtained that the $s$
quark has the softer spectrum than the $\bar s$, corresponding to a negative
$S^-$. Using a light-cone two-body wave function model applied to
$\Lambda K$ fluctuations \cite{Brodsky:1996hc}, a positive result
similar to ours was obtained in
\cite{Ding:2004ht}, and in \cite{Ding:2004dv} an even larger reduction
of the anomaly than ours was found using an effective chiral quark
model. Unfortunately, none of these studies provide comparisons with
the measured strange sea, making the significance of their results
difficult to assess.

The experimental situation is at present unclear. In \cite{Barone:1999yv}, a positive $S^-$ was favored based on several earlier experiments, whereas  NuTeV \cite{Zeller:2002du,Mason:2004yf} obtains the opposite sign based on their own data. However, the global analysis in \cite{Olness:2003wz} of the $s-\bar s$ asymmetry, including the NuTeV data as well as the CCFR data and using a very general functional form for $s^-(x)$, gives a best fit value for the asymmetry $S^-$ of the same magnitude and sign as ours.

It is clear that an asymmetry in the strange quark sea of the nucleon
arises naturally in models where non-perturbative sea quark
distributions originate from hadronic fluctuations of the nucleon. Our
model, which reproduces the measured strange sea, reduces the NuTeV
anomaly to a level which does not give a significant indication of
physics beyond the Standard Model.

%%%%%%%%%%%%%%%%%%%%%%%%%%%%%%%%%%%%%%%%%%%%%%%%%%%%
\subsection{Comparison to standard parameterizations}
\label{sec:cteq}

Let us first note that it is a
non-trivial result that the many different data sets shown above can
be well described by a model with only the eight parameters in
eq.~(\ref{eq:params}). The model provides all the different parton
distributions in $x$ at $Q^2_0$, \ie\ $u_{v+sea}$, $d_{v+sea}$, $g$,
$\bar{u}$, $\bar{d}$, $s$, $\bar{s}$. This is achieved with only four
shape parameters (the $\sigma$'s) and three normalization parameters
(the $\alpha_{MB}$'s), \ie\ significantly fewer shape and
normalization parameters than distributions.

The simplest possible conventional parameterization based on counting
rules \cite{counting-rules}, $xf_i(x)=A_i x(1-x)^{a_i}$, cannot be as
economic in the number of parameters and still allow different shapes
of different sea distributions. Considering only valence and light sea
distributions and neglecting the strange sea, we were able to get
reasonable fits of the same data sets with eight free parameters (when
applying the constraints in
eq.~(\ref{eq:flavoursumrule},\ref{eq:mom-sumrule})). However, the
fitted values for the exponents $a_i$ for the sea quarks and the gluon
turn out more than twice as large as expected from the counting rules
(using the expected values gives a poor description of the data). With
our model we obtain a generally better description of the data, which
is achieved with fewer parameters. Moreover, a physically motivated
model whose parameters have a physical interpretation can give new
insights about the unknown non-perturbative QCD dynamics embodied in
the parton distributions.

The standard parameterizations of parton densities, such as CTEQ~\cite{CTEQ} and MRST~\cite{MRST}, have many more parameters (typically $20$ for the $x$-shapes and additional ones for normalizations) and provide high-quality fits to large sets of different kinds of data. 
It is, therefore, interesting to compare the parton densities resulting
from our model with such standard parameterizations. 

\begin{figure}[t]
\begin{center}
\includegraphics*[width=16cm]{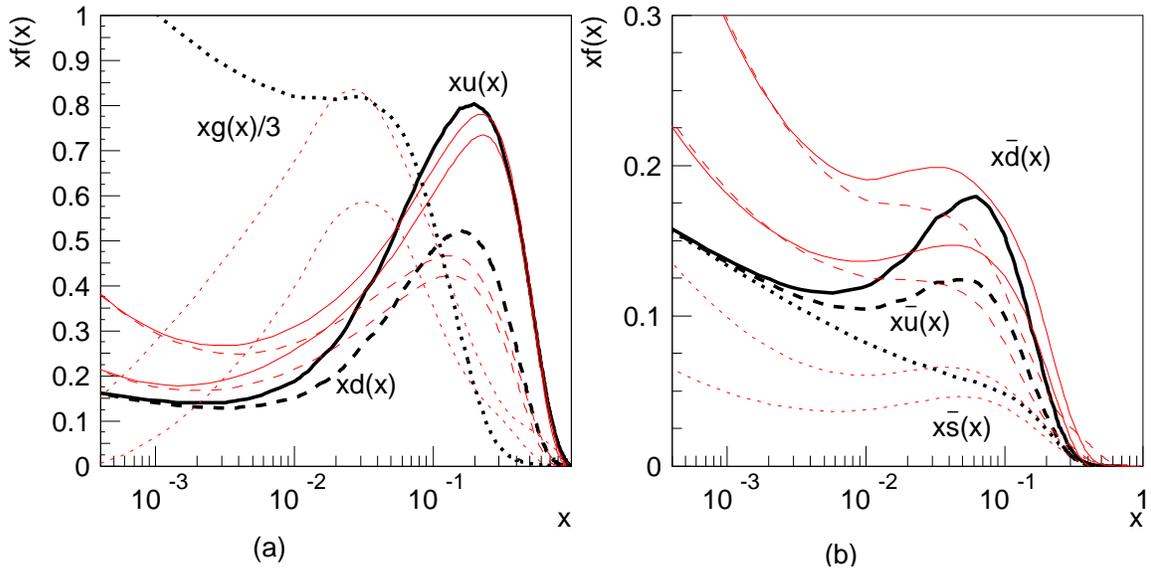}
\caption{\label{fig:pdf-cteq} Comparison between parton densities
at $Q^2=1.3\GeV^2$ from our model (thick lines) and the CTEQ6M
uncertainty band \cite{CTEQ} (plotted using the Durham HEPDATA server
\cite{durham}). Note the different scales on the axes of (a) and (b)
and that the gluon distributions are divided by 3 to fit in the plot.}
\end{center}
\end{figure}

In Fig.~\ref{fig:pdf-cteq} we show our parton densities together with the
uncertainty band of the CTEQ6M distributions \cite{CTEQ} at the low scale
$Q^2=1.3\GeV^2$. Several features of the plot are striking:
\begin{enumerate}
\item For large and intermediate $x$, our valence quark as well as sea
quark distributions from the model agree quite well with those from
CTEQ6M.
\item Our gluon distribution is slightly lower at large $x$ and much larger at small $x$.
\item For small $x\lesssim 10^{-2}$ the quark and anti-quark
distributions of CTEQ are clearly higher, and growing faster as
$x\to0$.
\end{enumerate}

The last two points are connected; if there is a deficit in low-$x$
sea quarks of non-perturbative origin in our model, this must be
compensated by perturbatively generated sea quarks from $g\to
q\bar{q}$ in the PQCD evolution in order to fit HERA $F_2$ data. This
requires a large gluon density at small $x$ and a low starting scale
$Q_0^2$, which combined with the momentum sum rule
eq.~(\ref{eq:mom-sumrule}) gives the low width $\sigma_g$ of the gluon
distribution. 
If instead of using next-to-leading order QCD evolution equations, we use only leading order ones, then the evolution is slower which makes the fit to data prefer an even lower starting scale $Q^2_0$ and a still more peaked gluon distribution at small $x$, \ie\ a smaller $\sigma_g$, in order to develop a large enough $q\bar{q}$ sea at low $x$. Thus, these two model parameters depend on the order used in the perturbative evolution, whereas the other parameters are not affected much. Furthermore, a less good fit to data is obtained with the leading order evolution.

The Gaussian momentum fluctuations in our model leads to parton number
densities that go to a constant as $x\to 0$ and hence momentum
densities going to zero as $x\to 0$. The increase of the momentum
distributions at $x\to 0$ seen in Fig.~\ref{fig:pdf-cteq} is thus
caused by PQCD evolution from the starting scale $Q_0^2$, which is
forced to be very low. We cannot, however, get a sea quark
distribution rising as fast as the $x^{-0.3}$-shape obtained by CTEQ
(as seen in Fig.~\ref{fig:pdf-cteq}b) at their $Q_0=1.3\GeV$, without
an accompanying rise in the $xg(x)$ density. This indicates that,
besides meson-baryon fluctuations, there is some additional source of
$q\bar q$-pairs. One possibility, which seems quite natural, would be
a non-perturbative ``gluon splitting'' process similar to the one in
PQCD evolution.  However, the precise form and effect of such a
process is not known. Such a contribution wouldn't have any large
effect on the asymmetries studied in this work, since the asymmetry
data are found at larger $x$ and $Q^2$. Therefore we have chosen not
to complicate our model by further assumptions and parameters to add
such a conceivable component.

%%%%%%%%%%%%%%%%%%%%%
\section{Conclusions}
%%%%%%%%%%%%%%%%%%%%%

We have presented a model based on Gaussian momentum fluctuations and
hadronic fluctuations of baryons into baryon-meson pairs. The
resulting parton momentum distributions describes a wealth of data
surprisingly well in view of the simplicity of the model. In this
study, we have focused on different asymmetries of quark distributions
in the proton; most notably the $u$--$d$ asymmetry obtained from the
$W^\pm$ forward-backward asymmetry at the Tevatron and the
$\bar{d}$--$\bar{u}$ asymmetry measured in $pp$ and $pd$ scattering.
From the phenomenological success of the model we can draw some
overall conclusions:
\begin{itemize}
\item The Gaussian, based on the law of large numbers, used for the momentum distribution may justify a statistical description of the non-perturbative forces confining partons in hadrons.
\item Asymmetries in the quark sea can be explained by 
hadronic fluctuations, such as $|p\rangle = \alpha_0|p_0\rangle + \alpha_{p\pi}|p\pi^0\rangle + \alpha_{n\pi}|n\pi^+\rangle + \ldots + \alpha_{\Lambda K}|\Lambda K^+\rangle + \ldots $
\end{itemize}

From the details of our study, we can make some further assertions:
\begin{itemize}
\item The difference in shape between the $u$ and $d$ valence
distributions of the proton corresponds to a difference in the widths
of their momentum distributions, which may be interpreted as a difference in the spacial region available due to Pauli blocking.
\item Comparison with the experimental result for the distribution of 
$\bar d(x)-\bar u(x)$ indicates that a larger effective mass is needed
for the pion in the $|N\pi\rangle$ states, which may be interpreted as an effective way of also including fluctuations with heavier mesons.
\item Besides fluctuations into hadronic states, some other
mechanism seems to be needed to explain the rise in the sea
$q\bar q$-distributions as $x\to 0$, used in standard
parameterizations of parton distributions.
\item An asymmetry between the strange and anti-strange nucleon sea
is unavoidable, although the magnitude of the asymmetry  $S^-=\int_0^1dx(xs(x)-x\bar s(x))$ cannot be
stated more precisely than $0.0010\leq S^-\leq 0.0023$. This reduces the NuTeV anomaly from three to about two standard deviations, leaving no significant indication for physics beyond the standard model.
\end{itemize}
  
Although our model invoke similar basic hadronic fluctuations as in
meson cloud models, it differs in important aspects, primarily
concerning the basis for the momentum distributions of both the
hadrons in the fluctuations and the partons in the hadrons. As discussed above, our model is also more economical in terms of using
fewer parameters than even the simplest parameterizations of parton densities.
Our model provides the parton distributions for all
light quarks, as well as the strange sea and gluons, with altogether
only four parameters giving the $x$-shapes and three parameters for
normalizations.

Finally, we note that this model could be used to develop a library of parton density functions for more general usage. So far, we have not done this since our intention has not been high precision functions to be folded with parton level cross-sections in practical calculations. Instead, our primary aim has been to gain understanding of the non-perturbative QCD dynamics of the bound state nucleon that is embodied in the parton density functions at the low scale $Q^2_0$, which are otherwise just parameterized.

{\bf Acknowledgments:} This research was supported by the Swedish Research Council. 

%%%%%%%%%%%%%%%%%%%%%%%%%%%

\end{document}